\documentclass[aps,prl,twocolumn,groupedaddress,showpacs,%
amsmath,preprintnumbers]{revtex4}
\usepackage{hyperref}
\def\half{{\scriptstyle\frac{1}{2}} }
\def\BTZ{{\scriptscriptstyle\mathit{BTZ}}}
\begin{document}

\preprint{UCD-03-11} 
\preprint{hep-th/0311090}
\title{Non-Quasinormal Modes and Black Hole Physics}
\author{Danny Birmingham}
\email[]{birm@itp.stanford.edu}
\altaffiliation{on leave from Dept.\ of Mathematical Physics,
University College Dublin, Ireland}
\author{S.\ Carlip}
\email[]{carlip@dirac.ucdavis.edu}
\affiliation{Department of Physics,
University of California, Davis, CA 95616}

\date{\today}

\begin{abstract}
The near-horizon geometry of a large class of extremal and 
near-extremal black holes in string and M theory contains three-dimensional 
asymptotically anti-de Sitter space.  Motivated by this structure, we are led 
naturally to a discrete set of complex frequencies defined in terms of the 
monodromy at the inner and outer horizons of the black hole.  We show 
that the correspondence principle, whereby the real part of these 
``non-quasinormal frequencies'' is identified with certain fundamental 
quanta, leads directly to the correct quantum behavior of the near-horizon 
Virasoro algebra, and thus the black hole entropy. Remarkably, for the  
rotating black hole in five dimensions we also reproduce the fractionization 
of conformal weights predicted in string theory.
\end{abstract}

 
\maketitle
\section{Introduction}

It is generally expected that a thorough understanding of the quantum 
mechanical nature of black holes will help pave the way towards a fully 
consistent theory of quantum gravity.  In this context, one recent effort  
has focused on applying the correspondence principle to black hole 
physics. The central idea \cite{Bekenstein, Hod} is to identify the real part
of the classical quasinormal frequencies of gravitational perturbations
with the elementary quanta of black hole mass and angular momentum. 
Interest in such a proposal increased with the observation \cite{Dreyer} 
that such a correspondence can correctly fix the Immirzi parameter
\cite{Immirzi}, an undetermined prefactor in the area operator of loop 
quantum gravity.  However, a general understanding of the universality 
of such an approach is still lacking. In particular, the applicability to 
charged and rotating black holes remains unclear; see for example 
\cite{Motl,Hod2,Berti1,Berti2,Hod3,Oppenheim,Musiri}.

In \cite{BCC}, we studied the correspondence principle as it applies to 
the $(2+1)$-dimensional black hole of Ba\~{n}ados, Teitelboim, and 
Zanelli ({\rm BTZ}) \cite{BTZ}.  The quantum mechanics of the {\rm BTZ} 
black hole is characterized by a Virasoro algebra. By identifying the real 
part of the quasinormal modes with the fundamental quanta of black 
hole mass and angular momentum, we found that an elementary excitation 
corresponds exactly to a correctly quantized shift of the Virasoro generators 
$L_{0}$ and $\bar{L}_{0}$ of this conformal algebra. Furthermore, we 
showed that by applying the correspondence principle to the mass and 
angular momentum parameters of point particle spacetimes and their 
analogs in Liouville theory \cite{Chen}, one can also account for transitions 
among ground states whose  values of $L_{0}$ and $\bar{L}_{0}$ do not 
differ by integers.

In this paper, we consider general classes of four- and five-dimensional
near-extremal, charged, rotating black holes that arise in string and 
M theory \cite{Cvetic4d,Horowitz,Cvetic5d}.  The  near-extremal 
Reissner-Nordstr\"{o}m black hole occurs as a special case.
Although these black holes are asymptotically flat, they have the special 
feature that the geometry in the near-horizon region contains that of 
the {\rm BTZ} black hole \cite{Hyun,Sken}.  In particular, it has been 
shown that the associated near-horizon conformal symmetry leads directly 
to a microscopic understanding of the Bekenstein-Hawking entropy 
\cite{Sken,Larsen4d,Larsen5d,Larsen4drot}.  Our aim is to understand 
how the correspondence principle might be applied to these higher-%
dimensional black holes to capture the correct quantization of this
symmetry.

Conventionally, quasinormal mode boundary conditions are stated in 
terms of behavior of perturbations at the horizon and infinity. This 
has always been somewhat mysterious, since it is not clear why black hole 
quantization should care about infinity.   Here,  we first show that for 
the {\rm BTZ} black hole, these boundary conditions can be recast in terms 
of monodromy conditions at the inner and outer horizons \cite{Siopsis}. 
Armed with this knowledge, we are led naturally 
to define a set of ``non-quasinormal modes'' for the higher-dimensional
black holes involving only these monodromies. The correspondence 
principle leads directly to the correct quantization of the near-horizon 
Virasoro generators.  Remarkably, for the rotating black hole in five 
dimensions, the correspondence principle also leads to the correct 
fractionization of conformal weights, in exact agreement with the 
predictions of a microscopic D-brane analysis \cite{Vafa1,Vafa2,Susskind}.

\section{Horizon Monodromies}

For black holes in asymptotically anti-de Sitter spacetimes, the quasinormal 
modes are defined as solutions that are ingoing at the horizon and either 
vanish \cite{Hubeny} or have vanishing flux \cite{BSS} at infinity.  Let us 
first consider the quasinormal modes of the {\rm BTZ} black hole, which 
is a solution of the vacuum Einstein equations in three spacetime dimensions 
with negative cosmological constant $\Lambda = -1/l^2$. The black hole is 
parametrized by its ADM mass $M_{\BTZ}$ and angular momentum 
$J_{\BTZ}$, and has inner and outer horizons at $\rho_{\pm}$, with
\begin{equation}
8G_{3}M_{\BTZ} = \frac{\rho_{+}^{2} + \rho_{-}^{2}}{l^{2}}, \quad
8G_{3}J_{\BTZ} = \frac{2 \rho_{+}\rho_{-}}{l}.
\end{equation}
 
For simplicity, we consider the quasinormal modes of an 
infinitesimal \footnote{Finite perturbations may destabilize the inner 
horizon \cite{Ori}; for now, we evade this issue by considering only 
infinitesimal ``test'' fields to probe the black hole structure.} massless 
scalar perturbation $ \Phi$ satisfying $\nabla^{2}\Phi = 0$.  The solution 
can be written as
\begin{equation}
\Phi = R(\rho) 
\exp\{-i\omega_{\BTZ}t_{\BTZ} 
+ i k_{\BTZ}\phi_{\BTZ}\},
\end{equation}
where the radial function satisfies the hypergeometric equation with regular 
singular points at the inner and outer horizons and infinity \cite{DB,CardBTZ}.  
Near the outer horizon, the radial part of the ingoing solution 
takes the form
\begin{equation}
R(\rho_{+})  \sim (\rho^{2} - \rho_{+}^{2})^{-\frac{i}{8\pi}
    \left[\beta_{R}\left(\omega^{\BTZ} + \frac{k^{\BTZ}}{l}\right) 
+ \beta_{L}\left(\omega^{\BTZ} - \frac{k^{\BTZ}}{l}\right)\right]} ,
\label{BTZouter}
\end{equation}
with $\beta_{R,L} = 2 \pi l^{2}/({\rho_{+} \pm \rho_{-}})$; the  inverse
temperatures at the inner and outer horizons are $\beta_\pm = 
(\beta_R\pm\beta_L)/2 = -\Omega_\mp\beta_\mp l$, where 
$\Omega_\pm$ are the angular velocities of the inner and outer 
horizons.  Inspired by \cite{Motl,Siopsis}, we consider the monodromy 
of this solution---the change under a $2\pi$ rotation in the complex 
$\rho$ plane around the zero at $\rho=\rho_+$---which is given by
\begin{multline}
{\cal M}(\rho_{+}) = \\
\exp\left\{\frac{1}{4}\left[\beta_{R} 
   \left(\omega^{\BTZ} + \frac{k^{\BTZ}}{l}\right) + \beta_{L} 
   \left(\omega^{\BTZ} - \frac{k^{\BTZ}}{l}\right)\right ]\right\}.  
\end{multline}

The conventional way to determine the quasinormal modes is to continue this 
solution to infinity and demand that it satisfy Dirichlet boundary conditions. 
We employ an alternative procedure.  We first continue the ingoing solution 
to the inner horizon, where it becomes a particular linear combination of the 
two independent solutions at $\rho_{-}$, with behaviors that follow from
standard properties of hypergeometric functions:
\begin{equation}
R^{\pm}(\rho_{-})  \sim (\rho^{2} - \rho_{-}^{2})^{\pm\frac{i}{8\pi}
   \left[ \beta_{R}\left(\omega^{\BTZ} + \frac{k^{\BTZ}}{l}\right) 
   - \beta_{L}\left(\omega^{\BTZ} - \frac{k^{\BTZ}}{l}\right)\right]}.
\label{BTZinner}
\end{equation}
The monodromies of these solutions around $\rho=\rho_-$ are
\begin{multline}
{\cal M}^{\pm}(\rho_{-}) =\\
\exp\left\{\pm\frac{1}{4}\left[ 
   \beta_{R}\left(\omega^{\BTZ} + \frac{k^{\BTZ}}{l}\right) 
   - \beta_{L}\left(\omega^{\BTZ} - \frac{k^{\BTZ}}{l}\right) 
   \right]\right\}.
\end{multline}
We now demand that the product of monodromies at the inner and 
outer horizons be trivial, ${\cal M} (\rho_{+}){\cal M}(\rho_{-}) = 1$.
It may be checked that this condition is satisfied  
if either ${\cal M} (\rho_{+}){\cal M}^{+}(\rho_{-}) = 1$
or ${\cal M} (\rho_{+}){\cal M}^{-}(\rho_{-}) = 1$, leading to 
frequencies (for ${\rm Im}\;\omega^{\BTZ} < 0)$
\begin{equation}
\omega_{L}^{\BTZ} = \frac{k_{L}^{\BTZ}}{l} 
- \frac{4 \pi i}{\beta_{L}}(n+1), \quad
\omega_{R}^{\BTZ} = \frac{k_{R}^{\BTZ}}{l} 
- \frac{4 \pi i}{\beta_{R}}(n+1),
\label{QNM}
\end{equation}
where the mode number $n\in{\bf N}$, and $k_{L}, k_{R} \in {\bf Z}$. These 
frequencies are immediately recognized as the quasinormal modes of the 
${\rm BTZ}$ black hole \cite{DB,CardBTZ}.  While the monodromy condition involves 
the behavior of the perturbation only in the region between the inner and 
outer horizon, it is nevertheless equivalent to the conventional quasinormal 
mode boundary conditions involving behavior outside the outer horizon. This 
reinterpretation has been discussed in the static case in \cite{Siopsis}.

\section{Charged 4-Dimensional Black Holes}

We now consider a class of four-dimensional charged black holes arising
in M theory \cite{Cvetic4d, Horowitz}.  For our purposes, it is 
convenient to reinterpret these as five-dimensional black string solutions 
by implementing a boost along a fifth direction \cite{Larsen4d}.  The 
line element then takes the form
\begin{align}
ds^{2} = &(H_{1}H_{2}H_{3})^{-\frac{1}{3}}[-fd\tilde{t}^{2} + d\tilde{x}^{2}]\nonumber\\
&+ (H_{1}H_{2}H_{3})^{\frac{2}{3}}[f^{-1}dr^{2} + r^{2}d\Omega_{2}^{2}],
\end{align}
where $f = 1 - {r_{0}}/{r}$, $\tilde{t} = t\cosh \delta_{0} - x\sinh\delta_{0}$, 
and $\tilde{x} = - t\sinh\delta_{0} + x\cosh\delta_{0}$. The harmonic 
functions $H_{i}$ are $H_{i} = 1 + \frac{r_{0}\sinh^{2}\delta_{i}}{r}$, and 
the physical charges are $Q_{i} = \half r_{0}\sinh 2\delta_{i}$.  The 
general solution depends on five parameters $r_{0}$, $\delta_{0}$, 
$\delta_{1}$, $\delta_{2}$, $\delta_{3}$; the case  $\delta_{0} = 
\delta_{1} = \delta_{2} = \delta_{3}$ is the standard Reissner-Nordstr\"{o}m  
black hole.

We wish to study a perturbation of the black string by a massless scalar 
field $\Phi$ satisfying $\nabla^{2}\Phi = 0$.  Using the ansatz 
$\Phi = e^{-i\tilde{\omega}\tilde{t} + i \tilde{k}\tilde{x}}R(r)Y(\theta,\phi)$, 
the full five-dimensional wave equation can be written down explicitly. 
For our discussion, we need only consider the leading terms close to the 
horizons, namely \cite{Cvetic}
\begin{multline}
\frac{d}{dx}\left[\left(x^{2} - \frac{1}{4}\right)\frac{dR}{dx}\right] \\
+ \frac{1}{x - \half}(\tilde{\omega}r_{0} \cosh \delta_{1} \cosh\delta_{2} 
\cosh\delta_{3})^{2}R  \\
-\frac{1}{x + \half}(\tilde{k}r_{0} \sinh \delta_{1} \sinh\delta_{2} 
\sinh\delta_{3})^{2}R = 0,
\end{multline}
where we have introduced the coordinate $x = \frac{r}{r_{0}} - \frac{1}{2}$.

As usual, we first select a solution that is ingoing near the outer horizon 
$x = \frac{1}{2}$.  This has the form
\begin{equation}
R  \sim \left( x - \half\right)^%
   {-i\tilde{\omega} r_{0}\cosh \delta_{1}\cosh\delta_{2}\cosh\delta_{3}}.
\label{4douter}
\end{equation}
The general solution near the inner horizon $x = -\frac{1}{2}$ is
\begin{align}
R \sim A_{+} &\left(x + \half\right)^%
    {i\tilde{k} r_{0} \sinh\delta_{1} \sinh \delta_{2}\sinh \delta_{3}} \nonumber\\
    &+ A_{-} \left(x + \half\right)^%
    {-i\tilde{k} r_{0} \sinh\delta_{1} \sinh \delta_{2}\sinh \delta_{3}}.
\label{4dinner}
\end{align}
In the dilute gas regime $\delta_{1}, \delta_{2}, \delta_{3} \gg 1$, the demand
that the product of the monodromies at the two horizons be trivial then
determines the frequencies (with ${\rm Im}\; \omega < 0$)
\begin{equation}
\omega \pm k = -  \frac{8 (n +1) i}{r_{0}}
   \exp\{-(\delta_{1} + \delta_{2} + \delta_{3}) \pm  \delta_{0}\},
\label{nonqnm}
\end{equation}
where $ \omega = \tilde{\omega}\cosh \delta_{0} + \tilde{k} \sinh \delta_{0}$, 
$k = \tilde{\omega}\sinh\delta_{0} + \tilde{k} \cosh\delta_{0}$, and
$n \in {\bf N}$.  Note that the dilute gas condition implies that the black 
hole is necessarily near-extremal.

To make contact with our previous discussion, recall that the geometry in 
the near-horizon limit takes the form ${\rm BTZ} \times S^{2}$, with the 
mass and charge of the black string reinterpreted in terms of the mass and 
angular momentum of the {\rm BTZ} black hole \cite{Hyun,Sken,Larsen4d}. 
Specifically, one finds the identification of coordinates \cite{Larsen4d}
\begin{equation}
t_{\BTZ} = \frac{tl}{R_{x}},\quad \phi_{\BTZ} = \frac{x}{R_{x}},\quad 
\rho^{2} = \frac{2R_{x}^{2}}{l}(r + r_{0}\sinh^{2}{\delta_{0}}),
\label{4dident1}
\end{equation}
with effective three-dimensional mass and angular momentum parameters  
\begin{equation}
8G_{3}M_{\BTZ} = \frac{2r_{0}R_{x}^{2}}{l^{3}}\cosh{2\delta_{0}},\ \
8G_{3}J_{\BTZ} = \frac{2r_{0}R_{x}^{2}}{l^{2}}\sinh{2\delta_{0}},
\label{4dident2}
\end{equation}
where $R_{x}$ is the radius of the compact $x$-direction.  The   
three-dimensional cosmological constant is $\Lambda = -1/l^2$, with
$l = 2(Q_{1}Q_{2}Q_{3})^{\frac{1}{3}}$, and the gravitational coupling 
is  $G_{3} = 2R_xG_{4}/l^{2}$.  Using the identifications (\ref{4dident1}) 
and (\ref{4dident2}) and the dilute gas condition $\delta_{1},\delta_{2}, 
\delta_{3}\gg 1$, one can check that the frequencies (\ref{nonqnm}) agree 
precisely with (\ref{QNM}).

It is important to stress that the black string solution is asymptotically flat, and 
conventional quasinormal mode boundary conditions would demand an 
outgoing solution at infinity.  Our monodromy condition, however, requires that
either $A_{+} = 0$ or $A_{-} = 0$, and is incompatible with such boundary 
conditions.  We shall call our novel set of frequencies ``non-quasinormal modes."

(Further ``stringy'' information can be obtained from purely ingoing modes 
used to compute greybody factors \cite{AGMOO}, but these require 
different boundary conditions, and in particular lead to real, nonquantized
frequencies.)

We can now apply the correspondence principle, identifying the real part of 
these frequencies with certain fundamental quanta.  The key is the 
presence of a near-horizon conformal symmetry.  The associated Virasoro 
generators can be expressed in terms of the effective three-dimensional mass 
and angular momentum (\ref{4dident2}) as \cite{Larsen4d}
\begin{align}
L_{0} &= \frac{1}{2}(M_{\BTZ}l + J_{\BTZ}) + \frac{l}{16 G_{3}}  
  = \frac{R_x}{2}(E + P) + \frac{l}{16 G_{3}},\nonumber\\
\bar{L}_{0} &= \frac{1}{2}(M_{\BTZ}l - J_{\BTZ}) + \frac{l}{16 G_{3}} 
  = \frac{R_x}{2}(E - P) + \frac{l}{16 G_{3}},
\end{align}
where $E = \frac{r_{0}}{8G_{4}}\cosh 2\delta_{0}$ and 
$P = \frac{r_{0}}{8G_{4}}\sinh 2\delta_{0}$ are the energy and momentum 
of the black string flowing in the $x$-direction.  Following \cite{BCC}, we identify 
the real part of the non-quasinormal modes with the fundamental quanta
of $M_{\BTZ}$ and $J_{\BTZ}$, via
\begin{eqnarray}
\Delta M_{\BTZ} &=& \omega_{L}^{\BTZ} + \omega_{R}^{\BTZ} 
  = \frac{k_{L}^{\BTZ}}{l} + \frac{k_{R}^{\BTZ}}{l},\nonumber\\
\Delta \left(\frac{J_{\BTZ}}{l}\right) 
  &=& \omega_{L}^{\BTZ} - \omega_{R}^{\BTZ}
  = \frac{k_{L}^{\BTZ}}{l} - \frac{k_{R}^{\BTZ}}{l}.
\label{corr}
\end{eqnarray}
This leads directly to quantization of the Virasoro operators, 
$\Delta L_{0} = k_{L}^{\BTZ}$ and $\Delta \bar{L}_{0} = k_{R}^{\BTZ}$. 
As a result, we obtain the correct quantization of the energy and momentum 
of the black string,
\begin{equation}
\Delta E = \frac{1}{R_x}(k_{L}^{\BTZ} + k_{R}^{\BTZ}),\quad
\Delta P = \frac{1}{R_x}(k_{L}^{\BTZ} - k_{R}^{\BTZ}).
\end{equation}

\section{Charged Rotating 5-Dimensional Black Holes}

We next consider a class of five-dimensional charged rotating black holes 
that arise in M theory \cite{Cvetic5d}. Our aim is to explore the new features  
coming from the rotating nature of the black hole.  As above, these 
configurations can be reinterpreted as black strings, now in six dimensions. 
The general solution depends on six parameters \cite{Larsen5d}---a mass $m$, 
charges $Q_{1} = m\sinh 2 \delta_{1}$, $Q_{2} = m\sinh 2 \delta_{2}$, angular 
momentum parameters $l_{1}$, $l_{2}$, and a boost parameter $\delta_{0}$
along the compact string direction $y$.  The system has a microscopic 
interpretation in terms of D-branes, charged with respect to $Q_{1}$ and 
$Q_{2}$: it corresponds to a collection of $n_{1}$ D1-branes and $n_{2}$ 
D5-branes, with $n_1 = \frac{\pi}{4}\frac{g\alpha' Q_1}{G_5R_y}$ and 
$n_2=\frac{Q_2}{g\alpha'}$, where $g$ and $\alpha'$ are the string 
coupling and Regge slope and $R_{y}$ is the radius in the $y$ direction.

The near-horizon geometry is of the form ${\rm BTZ} \times S^{3}$, with 
effective three-dimensional mass, angular momentum, gravitational and 
cosmological constants \cite{Larsen5d}
\begin{align}
8G_{3}M_{\BTZ} &= 
   \frac{R_{y}^{2}}{l^{4}}\left[
   (2m - l_{1}^{2} - l_{2}^{2})\cosh 2\delta_{0} + 2 l_{1}l_{2}\sinh 2\delta_{0}\right],
   \nonumber\\
8G_{3}J_{\BTZ} &=
   \frac{R_{y}^{2}}{l^{3}}\left[
   (2m - l_{1}^{2} - l_{2}^{2})\sinh 2\delta_{0} + 2 l_{1}l_{2}\cosh 2\delta_{0}\right]
   \nonumber\\
l & =(Q_{1}Q_{2})^{1/4}, \quad G_{3} = R_{y}G_{5}/\pi l^{3} .
\label{5dident}
\end{align}
As before, we can study the full six-dimensional wave equation to 
look for the solutions near the inner and outer horizons.  Using the 
identifications (\ref{5dident}), we again find that the non-quasinormal 
frequencies are given by (\ref{QNM}).

Using (\ref{5dident}), the Virasoro generators take the form \cite{Larsen5d}
\begin{eqnarray}
L_{0} &=& \frac{R_{y}}{2}(E + P) - \frac{1}{n_{1}n_{2}}j_{L}^2 + \frac{n_{1}n_{2}}{4},
   \nonumber\\
\bar{L}_{0} &=& \frac{R_{y}}{2}(E - P) - \frac{1}{n_{1}n_{2}}j_{R}^2 + \frac{n_{1}n_{2}}{4},
\label{5dVir}
\end{eqnarray}
where the energy $E$, momentum $P$, and angular momenta $j_{L,R}$ of the black 
string are  
\begin{align}
&E = \frac{\pi}{4 G_{5}}m \cosh 2 \delta_{0},\quad
P = \frac{\pi}{4 G_{5}}m \sinh 2 \delta_{0},\quad\nonumber\\
&j_{L,R} = \frac{\pi}{16G_{5}}(l_{1} \mp l_{2})m e^{\pm\delta_{0} + \delta_{1} + \delta_{2}},
\end{align}
and the Virasoro algebra has central charge $c= 6n_{1}n_{2}$.  

In the absence of angular momentum, the correspondence principle (\ref{corr})
again leads to the correct integer shifts in the energy and momentum of the string.
To understand the angular momentum contribution in (\ref{5dVir}), we recall that
the microscopic states of the {\rm BTZ} black hole can be described by Liouville
theory \cite{Coussaert,Bautier}, and that Liouville theory contains additional states 
whose values of $L_{0}$ and $\bar{L}_{0}$ agree precisely with those of point particle 
spacetimes in $2+1$ dimensions \cite{Chen}.  The Virasoro generators in this case 
are given by
\begin{eqnarray}
L_{0} &=& -\frac{l}{16 G_{3}}\left[1 - 4G_{3}\left( m + \frac{j}{l}\right)\right]^{2} 
+ \frac{l}{16 G_{3}},\nonumber\\
\bar{L}_{0} &=& -\frac{l}{16 G_{3}}\left[1 - 4G_{3}\left( m - \frac{j}{l}\right)\right]^{2} 
+ \frac{l}{16 G_{3}}.
\end{eqnarray}
By applying the correspondence principle to $m$ and $j$, via $\Delta (m + \frac{j}{l}) 
= 2 k_{L}^{\BTZ}/l$ and $\Delta (m - \frac{j}{l}) = 2 k_{R}^{\BTZ}/l$, one is led 
to quantization of the form \cite{BCC}
\begin{equation}
\Delta L_{0} = k_{L}^{\BTZ} - \frac{1}{n_{1}n_{2}}(k_{L}^{\BTZ})^{2},\quad
\Delta \bar{L}_{0} = k_{R}^{\BTZ} - \frac{1}{n_{1}n_{2}}(k_{R}^{\BTZ})^{2}.
\end{equation}
Remarkably, this fractionization of conformal weights is in precise agreement
with the results of the D-brane analysis \cite{Vafa1,Vafa2,Susskind}.  A similar 
fractionization occurs in the four-dimensional rotating solution \cite{Larsen4drot}.

\section{Discussion}

In the absence of a complete quantum theory of gravity, it is natural to
speculate that classical frequencies might be related to quantum energy 
levels.  Our results have confirmed this intuition for a large class of 
``stringy'' black holes, demonstrating that many features of 
quantization---even those as special as fractionization of conformal
weights---fit such a correspondence principle.  On the other hand, the
modes we have considered are {\em not\/} the usual quasinormal modes,
but rather a new set picked out by behavior near the inner and outer
horizons.  In one sense, this is physically sensible: black hole quantization
ought to depend on properties at the horizon, not those at infinity, and 
many would find the symmetric appearance of the inner and outer horizons 
appealing.  On the other hand, our results further deepen the mystery of 
why ordinary quasinormal modes, and not our ``non-quasinormal modes,'' 
seem to be so closely related to the Immirzi parameter of loop quantum 
gravity.  There is clearly much more to be understood.

\section{Acknowledgments}
 D.B.\ is grateful to the Departments of Physics at UC Davis and 
University of the Pacific for hospitality.  S.C.\ was supported in part by U.S.\ 
Department of Energy grant DE-FG03-91ER40674.

\end{document}